\documentclass[review,12pt]{elsarticle}
\usepackage{amsmath, amsthm, amssymb}

\journal{Physics Letters A}

\begin{document}
\title{Answer to the comment on ``On the characteristic
	polynomial of an effective Hamiltonian''}
\author{Yong Zheng  \corref{correspondingauthor}}
\cortext[correspondingauthor]{Corresponding author}
\ead{zhengyongsc@sina.com}
\address{School of Physics and Electronics, Qiannan Normal University for Nationalities,
	Duyun 558000, China}

\begin{abstract}
	In a recent comment by Fern\'{a}ndez,  it has been argued that our solution method of an effective Hamiltonian based on the characteristic polynomial [Phys. Lett. A 443 (2022) 128215] had been developed several years earlier by Fried and Ezra [J. Chem. Phys. 90 (1989) 6378]. 
We show here several important differences between our treatment and the resummation method proposed previously by Fried and Ezra. 
\end{abstract}


\maketitle

\section{Introduction}

Recently, Fern\'{a}ndez has critically pointed out that our solution method of an effective Hamiltonian based on the characteristic polynomial had been developed several years earlier by Fried and Ezra in a clearer and more general way \cite{C}. The resummation method
of Fried and Ezra based on the reconstruction of the effective secular equation \cite{Fried}, which we were unaware of in our previous study \cite{Z}, does have some similarities with our method. However, we show here that Fern\'{a}ndez has ignored several important differences between such two  methods. 

First of all, Fern\'{a}ndez has ignored that: Unlike in Fried and Ezra's treatment, for our characteristic polynomial of the effective Hamiltonian $H_{\mathrm{eff}}(\lambda)$,
\begin{multline}\label{edetN}
	\det[E-H_{\mathrm{eff}}(\lambda)]=\prod_{n=1}^{N}[E- E_n^{_P}(\lambda )] =E^{N}-P_1(\lambda)E^{N-1}\\
	+\cdots+(-1)^{N-1}P_{N-1}(\lambda)E+(-1)^{N}P_{N}(\lambda) ,
\end{multline}
 the coefficients are expressed with the symmetric polynomials of $P$-space eigenvalues (reproduced here for the convenience of discussion),
\begin{subequations}
	\begin{align}
		&P_1(\lambda)=E_1^{_P}(\lambda )+E_2^{_P}(\lambda )+\cdots+E_N^{_P}(\lambda ), \tag{2.1} \label{P1}\\
		&P_2(\lambda)=\sum_{1\leq j_1<j_2\leq N}E_{j_1}^{_P}(\lambda )E_{j_2}^{_P}(\lambda ),\tag{2.2} \label{P2}\\
		& \quad \cdots \cdots  \notag\\
		&P\!_{N}(\lambda)=E_1^{_P}(\lambda )E_2^{_P}(\lambda )\cdots E_{N}^{_P}(\lambda
		). \tag{2.N} \label{PN}
	\end{align}
\end{subequations}

Introducing these symmetric polynomials and discussing their analytical property in complex plane of $\lambda$ are one of the key contents in our study. Actually, without such discussion, the better $\lambda$-expansion property of the characteristic polynomial $\det[E-H_{\mathrm{eff}}(\lambda)]$ (the so-called secular determinant) in Fried and Ezra's treatment can only be inferred from the effective Hamiltonian itself \cite{Fried}, which had been thought to be rapidly convergent in $\lambda$-expansion. However, as our study has shown, the analytical properties of $\det[E-H_{\mathrm{eff}}(\lambda)]$ and $H_{\mathrm{eff}}(\lambda)$ generally are different \cite{Z}.

Additionally, since the analytical property of the characteristic polynomial coefficients is unknown in Fried and Ezra's treatment, the $\lambda$-expansion can only be carried out by taking  $\det[E-H_{\mathrm{eff}}(\lambda)]$ as a whole, to obtain an  approximate characteristic polynomial in a form as, say, to the order of $ \lambda^{K} $,
\begin{equation}
\left\{\prod_{n=1}^{N}\left[E-\sum_{j=0}^{K} E_{n, j} \lambda^{j}\right]\right\}^{[K]}=E^{N}+\sum_{j=1}^{N} p_{j}(\lambda) E^{N-j},	\label{EF}
\end{equation}
where we have used the symbol $\{\cdots \}^{[K]}$ introduced by Fern\'{a}ndez, which means that one has to remove any term with $ \lambda^{k} $ if $ k > K $ and retain all the terms with $ \lambda^{k} $ if $ k \leq K $, to obtain a meaningful expansion\cite{C, Fried}; namely, the coefficients here $ p_{j}(\lambda)\equiv\sum_{k=1}^{K} c_{jk}\lambda^{k}$. This is equivalent to the expansion of all the $P_j(\lambda)$ in our Eq.~\eqref{edetN} to the same $\lambda$-order,  i.e, $P_j(\lambda)\approx (-1)^j p_{j}(\lambda)=(-1)^j \sum_{k=1}^{K} c_{jk}\lambda^{k}$. From this sense, Fried and Ezra's treatment seems to be equivalent to ours. However, their treatment generally can result in energy eigenvalues with an imaginary part, which are unphysical. This has already been noted by Fried and Ezra\cite{Fried}. 

The trouble of Fried and Ezra's treatment can be easily understood with the case of $N=2$, for which the energy  eigenvalues $ E_{1,2}(\lambda) $  can be solved with the approximate characteristic polynomial in Eq.~\eqref{EF}, 
\begin{equation}\label{eg2}
	E_{1,2}(\lambda) =\frac{-p_1(\lambda)\pm \sqrt{[p_1(\lambda)]^2-4p_2(\lambda)}}{2}.
\end{equation} 
Since we generally can not ensure that $ [p_1(\lambda)]^2 \geq 4p_2(\lambda) $, one can expect that energy eigenvalues with an imaginary part can be obtained when  
$[p_1(\lambda)]^2 < 4p_2(\lambda)$ (For example, when $K=1$, this $\lambda$-inequality is a quadratic one, and can be easily satisfied by certain $\lambda$ for a general Hamiltonian).

However, in our method, such trouble can be properly overcome. Our treatment is directly based on the $\lambda$-expansion of the coefficients $P_j(\lambda)$ via Eqs.~\eqref{P1}--\eqref{PN}, whose analytical property has been proven. Hence, we can carry out the expansion independently and properly retain terms to different orders of  $\lambda$ for different $P_j(\lambda)$, to avoid  problems such as $[p_1(\lambda)]^2 < 4p_2(\lambda)$ in the $N=2$ case above which possibly appear in Fried and Ezra's treatment. For example, for a given value of $\lambda$, we in principle can perform the expansion for each $P_j(\lambda)$  to some individual order, via requiring the additional correction caused by the next order within a certain precision range we set.  Whether the expansion is directly performed for $P_j(\lambda)$ or  $\det[E-H_{\mathrm{eff}}(\lambda)]$ is an important difference between the two treatments.

Fern\'{a}ndez also has made some misleading statements in the comment \cite{C}. For example, Fried an Ezra's method was described as a more general one just because they have not obviously introduced the total state space dimension $ \tilde{N}$ as we have done. However, Fern\'{a}ndez has not realized that we have not made any restrictions on such $ \tilde{N}$ in our study. Such $ \tilde{N}$ can be very large, since the dimension we really need to deal with is $N$. Actually, whether $N$ is large is more relevant. In fact, we have constructed the effective Hamiltonian  $ \bar{H}_{\mathrm{eff}}(\lambda) $ (see \cite{Z}) using the characteristic-polynomial coefficients $ P_n(\lambda) $.  Our  $ \bar{H}_{\mathrm{eff}}(\lambda) $ has been proven to hold a convergence radius same as $ P_n(\lambda) $ for $\lambda $-expansion. Hence, one can alternatively solve for the eigenvalues by diagonalizing $ \bar{H}_{\mathrm{eff}}(\lambda) $ in a matrix manner when $N$ is large.

Especially, Fern\'{a}ndez has argued that we had  overlooked the application of the method to estimate the exceptional
point in the complex  plane of  $ \lambda $ closest to origin. However, he has ignored that such application has already been performed by us in the last paragraph of Section 3 in \cite{Z}.

In conclusion, the two treatments of an  effective Hamiltonian based on the characteristic
polynomial, the one proposed by Fried and Ezra in \cite{Fried} and the one by us \cite{Z}, have important differences in specific  procedures.
The $\lambda$-expansion of the coefficients $P_j(\lambda)$ is important and should be performed carefully to avoid  the unphysical complex-energy-eigenvalue problem.
Fern\'{a}ndez has ignored these differences between the two treatments and some misleading statements have been made in the comment \cite{C}.

\end{document}